\documentclass[twocolumn,floatfix,aps,eqsecnum]{revtex4-2}
\usepackage{graphicx}
\usepackage{amsmath}
\usepackage{amssymb}
\usepackage{bm}
\usepackage{hyperref}

\usepackage{color}

\begin{document}

\title{Pair correlation function of the one-dimensional Riesz gas}
\author{C. W. J. Beenakker}
\affiliation{Instituut-Lorentz, Universiteit Leiden, P.O. Box 9506, 2300 RA Leiden, The Netherlands}
\date{December 2022}
\begin{abstract}
A method from random-matrix theory is used to calculate the pair correlation function of a one-dimensional gas of $N\gg 1$ classical particles with a power law repulsive interaction potential $u(x)\propto |x|^{-s}$ (a socalled Riesz gas). An integral formula for the covariance of single-particle operators is obtained which generalizes known results in the limits $s\rightarrow -1$ (Coulomb gas) and $s\rightarrow 0$ (log-gas). As an application, we calculate the variance of the center of mass of the Riesz gas, which has a universal large-$N$ limit that does not depend on the shape of the confining potential.
\end{abstract}
\maketitle

\section{Introduction}
\label{intro}

The one-dimensional (1D) Riesz gas \cite{Rie38,Rie49,Ser17,Lev22}, describes $N$ classical particles that move on a line (the $x$-axis) with a repulsive interaction potential $u(x)$ of the form
\begin{equation}
u(x)=\begin{cases}
\text{sign}(s) |x|^{-s}&\text{for}\;\;s>-2,\\
-\ln|x|&\text{for}\;\;s=0.
\end{cases}
\end{equation}
The particles are prevented from moving off to infinity by a confining potential $V(x)$.

The cases $s=0$ and $s=-1$ are also referred to as log-gas and Coulomb gas, respectively. The log-gas plays a central role in random-matrix theory (RMT) \cite{Forrester}. Experimentally, an interaction potential with an adjustable exponent $s\lesssim 1$ has been realized in a chain of trapped ion spins \cite{Zha17}.

Thermal averages $\langle\cdots\rangle$, at inverse temperature $\beta$, are defined with respect to the Gibbs measure
\begin{align}
&P(x_1,x_2,\ldots x_N)=Z^{-1}\nonumber\\
&\quad\times\exp\biggl(-\beta\biggl[J\sum_{i<j=1}^N u(x_i-x_j)+\sum_{i=1}^N V(x_i)\biggr]\biggr),
\end{align} 
where $Z$ normalizes the distribution to unity. The interaction strength is parameterized by $J>0$. The average density is
\begin{equation}
\rho(x)=\biggl\langle\sum_{i=1}^N\delta(x-x_i)\biggr\rangle,
\end{equation}
normalized to the particle number,
\begin{equation}
\int_{-\infty}^\infty\rho(x)\,dx=N.
\end{equation}

Much is known about the dependence of $\rho(x)$ on the range of the interaction \cite{Leb17,Har18,Aga19,Ket21}, in the three regimes $s> 1$ (short-range repulsion), $-1<s<1$ (weakly long-range repulsion), and $-2<s< -1$ (strongly long-range repulsion). In what follows we consider the second regime, $-1<s<1$, which includes the log-gas at $s=0$ and the Coulomb gas in the limit $s\to -1$.

For $N\gg 1$ the density has a compact support, which may consist of multiple disjunct intervals. Considering a single interval $(a,b)$, an end point may be $N$-independent, fixed by a hard wall, or it may be $N$-dependent, freely adjustable in a smooth potential. The density vanishes as a powerlaw at a free end point, while it diverges (with an integrable singularity) at a fixed end point. 

For example, in the case of a quadratic confinement $V(x)\propto x^2$, there are two free $N$-dependent end points at $b=-a\propto N^{1/(2+s)}$ and the density profile is \cite{Aga19}
\begin{equation}
\rho(x)\propto (b^2-x^2)^{(s+1)/2},\;\;|s|<1.
\end{equation}
The density profile becomes flat in the Coulomb gas limit, while the log-gas has the Wigner semicircle law \cite{Wig51}. Alternatively, if we set $V\equiv 0$ and confine the Riesz gas by a hard wall at $x=a$ and $x=b$, then the density diverges on approaching a fixed end point \cite{Ket21},
\begin{equation}
\rho(x)\propto(x-a)^{(s-1)/2}(b-x)^{(s-1)/2},\;\;|s|<1.\label{rhohardwall}
\end{equation}

Knowledge of the density allows us to calculate by integration the average $\langle F\rangle$ of a single-particle observable $F=\sum_{i=1}^N f(x_i)$ (also known as a ``linear statistic''). For $N\gg 1$ we can restrict the integral to the interval $(a,b)$,
\begin{equation}
\langle F\rangle=\int_a^b \rho(x)f(x)\,dx.
\end{equation}

Going beyond the first moment, Flack, Majumdar, and Schehr recently obtained \cite{Fla22} a strikingly simple formula for the large-$N$ limit of the variance of $F$ in the Coulomb case \cite{note1},
\begin{equation}
{\rm Var}\,F=\frac{1}{2\beta J}\int_a^b [df(x)/dx]^2\,dx\;\;\text{for}\;\;s=-1.\label{FMS}
\end{equation}
The corresponding formula in the RMT case $s=0$ is known \cite{Mehta,Bee96}, but results for other values of the interaction parameter $s$ are not known. The aim of this paper is to provide that information for the entire range $-1<s<1$. 

For that purpose one needs the connected pair correlation function
\begin{equation}
R(x,y)=\biggl\langle\sum_{i,j=1}^N\delta(x-x_i)\delta(y-x_j)\biggr\rangle-\rho(x)\rho(y),
\end{equation}
which gives the variance upon integration,
\begin{equation}
{\rm Var}\,F=\int_a^b dx\int_a^b dy\,R(x,y)f(x)f(y).
\end{equation}

In RMT there are basically two methods to compute $R(x,y)$. The method of orthogonal polynomials \cite{Meh60} applies to specific confining potentials (typically linear or quadratic) and then gives results for any $N$. The alternative method of functional derivatives \cite{Bee93a} (equivalently, the method of loop equations \cite{Amb90}) takes the large-$N$ limit, but then works generically for any form of confinement. Since the latter method does not assume a logarithmic repulsion, it is the method of choice in what follows.

\section{Pair correlation function}

For $N\gg 1$ the pair correlation function oscillates rapidly on the scale of the interparticle spacing $\delta x\simeq (b-a)/N$. These oscillations are irrelevant for the computation of the variance of an observable that varies smoothly on the scale of $\delta x$, so that in the large-$N$ limit it is sufficient to know the smoothed correlation function.

The method of functional derivatives starts from the exact representation
\begin{equation}
R(x,y)=-\frac{1}{\beta}\frac{\delta\rho(x)}{\delta V(y)}.\label{Kdef}
\end{equation}
The variation of the density is to be carried out at constant particle number,
\begin{equation}
\int_a^b\delta\rho(x)\,dx=0.\label{constraint}
\end{equation}
The integration interval $(a,b)$ is the support of the smoothed particle density $\rho(x)$ in the large-$N$ limit. (We assume that the confining potential produces a support in a single interval.)

In the regime $-1<s<1$ of a weakly long-ranged repulsion, and for $J\gtrsim 1/\beta$, variations in the smoothed density $\rho$ and in the confining potential $V$ are related by the condition of mechanical equilibrium \cite{Aga19,note2}, 
\begin{align}
J\int_a^b u(x-y)\delta\rho(y)\,dy+\delta V(x)={}&\text{constant},\nonumber\\
a<x<b.&\label{deltarhodeltaV}
\end{align}
Taking the derivative with respect to $x$ we have a singular integral equation,
\begin{align}
J(|s|+\delta_{0,s}){\cal P}\int_a^b dy\,\delta\rho(y)\frac{\text{sign}(x-y)}{|x-y|^{s+1}}={}&\frac{d}{dx}\delta V(x),\nonumber\\
a<x<b,&\label{inteq}
\end{align}
which we need to invert in order to obtain the functional derivative \eqref{Kdef}. (The symbol ${\cal P}$ indicates the principal value of the integral.)

For $-1<s<1$ the general solution to Eq.\ \eqref{inteq} is given by the Sonin inversion formula \cite{Son54,Sam93,Bul01} (see App.\ \ref{app_Sonin}),
\begin{widetext}
\begin{subequations}
\label{rhoVrelation}
\begin{align}
&J(|s|+\delta_{0,s})\delta\rho(x)=C_{\delta V}[(x-a)(b-x)]^{s_-}-C_1{\cal S}_{\delta V}(x),\\
&{\cal S}_{\delta V}(x)=-(x-a)^{s_+}\frac{d}{dx}\int_x^b dt\,\frac{(t-x)^{s_-}}{ (t-a)^{s}}\frac{d}{dt}\int_a^t dy\, (y-a)^{s_-}(t-y)^{s_+}\frac{d}{dy}\delta V(y),\\
&C_1=\frac{\sin(\pi s_+)\Gamma(s+1)}{\pi s_+\Gamma(s_+)^2},\;\;s_\pm=(s\pm 1)/2.
\end{align}
\end{subequations}
\end{widetext}
The coefficient $C_{\delta V}$ is fixed by the constraint \eqref{constraint},
\begin{subequations}
\begin{align}
&C_{\delta V}=\frac{C_1}{C_2}\int_a^b dx\,{\cal S}_{\delta V}(x),\\
&C_2=\int_a^b dx\,[(x-a)(b-x)]^{s_-}=\frac{(b-a)^s\sqrt{\pi}\,\Gamma(s_+)}{2^s\,\Gamma(1+s/2)}.
\end{align}
\end{subequations}
The function $\Gamma(x)$ is the usual gamma function.

The pair correlation function $R(x,y)$ is obtained from Eq.\ \eqref{Kdef} as a distribution, defined by its action on a test function $g(y)$ upon integration over $y$. Using the functional-derivative identities
\begin{subequations}
\begin{align}
&\int_a^b dy\,  \frac{\delta {\cal S}_{\delta V}(x)}{\delta V(y)}g(y) ={\cal S}_g(x),\\
&\int_a^b dy\,  \frac{\delta C_{\delta V}}{\delta V(y)}g(y) =\frac{C_1}{C_2}\int_a^b dx\,{\cal S}_g(x),
\end{align}
\end{subequations}
we find the following expression:
\begin{align}
&J\beta(|s|+\delta_{0,s})\,\int_a^b dy\, R(x,y)g(y)=C_1{\cal S}_g(x)\nonumber\\
&\quad -[(x-a)(b-x)]^{s_-}\frac{C_1}{C_2}\left(\int_a^b dx\,{\cal S}_{g}(x)\right).\label{Rgintegral}
\end{align}

\section{Covariance of single-particle observables}

Eq.\ \eqref{Rgintegral} provides a formula for the covariance of the two observables $F=\sum_{i=1}^N f(x_i)$ and $G=\sum_{i=1}^N g(x_i)$,
\begin{align}
{\rm CoVar}(F,G)={}&\langle FG\rangle-\langle F\rangle\langle G\rangle\nonumber\\
&=\int_a^b dx\int_a^b dy \,R(x,y)f(x)g(y).
\end{align}
The covariance is given by integrals over $f(x)$ and ${\cal S}_g(x)$,
\begin{align}
&J\beta(|s|+\delta_{0,s})\,{\rm CoVar}\,(F,G)=C_1\int_a^b dx\,f(x){\cal S}_g(x)\nonumber\\
&\quad -\frac{C_1}{C_2}\left(\int_a^b dx\,{\cal S}_{g}(x)\right)\int_a^b dx\,[(x-a)(b-x)]^{s_-}f(x).\label{CoVar}
\end{align}
The variance of a single observable then follows from ${\rm Var}\,F={\rm CoVar}(F,F)$.

For general functions $f,g$ the integrals in Eq.\ \eqref{CoVar} may be carried out numerically (see App.\ \ref{app_Sonin}). Closed-form expressions can be obtained for polynomial functions. It is convenient to shift the origin of the coordinate system so that $(a,b)\rightarrow (0,L)$. We define $X_p=\sum_i x_i^p$, $p\geq 1$, and obtain the covariance
\begin{widetext}
\begin{align}
J\beta(|s|+\delta_{0,s})\,{\rm CoVar}\,(X_p,X_q)={}&\frac{2 \pi    L^{p+q+s}\Gamma (s) }{ \Gamma (-p-s_-) \Gamma (p+s+1) \Gamma (-q-s_-) \Gamma (q+s+1) }\nonumber\\
&\times \frac{p q s\sin (\pi s_+)}{(p+q+s)\bigl(\cos [\pi  (p+q+s)]+\cos [\pi  (p-q)]\bigr)}.\label{CoVarXpXq}
\end{align}
For the variance this reduces to
\begin{align}
&J\beta(|s|+\delta_{0,s})\,{\rm Var}\,X_p=\frac{2\pi  L^{2 p+s} p^2 s\Gamma (s) \sin(\pi s_+) }{(2 p+s) \Gamma (-p-s_-)^2 \Gamma (p+s+1)^2\bigl(1+\cos[\pi(2p+s)]\bigr)}.\label{varresult}
\end{align}
\end{widetext}

We have checked that the general formula \eqref{CoVarXpXq} agrees with the known formulas in the Coulomb gas limit  \cite{Fla22} (see also App.\ \ref{app_Coulombgas}),
\begin{align}
&J\beta\,{\rm CoVar}\,(F,G)=\frac{1}{2}\int_0^L \frac{df(x)}{dx}\frac{dg(x)}{dx}\,dx,\;\;\text{for}\;\;s=-1,\nonumber\\
&\Rightarrow J\beta\,{\rm CoVar}\,(X_p,X_q)=\frac{p q L^{p+q-1}}{2 (p+q-1)},
\end{align}
and in the log-gas limit \cite{Bee94,Cun14},
\begin{widetext}
\begin{equation}
\begin{split}
&J\beta\,{\rm CoVar}\,(F,G)=\frac{1}{\pi^2}{\cal P}\int_0^L dx\int_0^L dy\,\frac{g(y)df(x)/dx }{y-x}\sqrt{\frac{x (L-x)}{y (L-y)}},\;\;\text{for}\;\;s=0,\\
&\Rightarrow J\beta\,{\rm CoVar}\,(X_p,X_q)=\frac{\pi   L^{p+q}}{(p+q) \Gamma \left(\frac{1}{2}-p\right) \Gamma (p) \Gamma \left(\frac{1}{2}-q\right) \Gamma (q)\cos (\pi  p) \cos (\pi  q)}.
\end{split}
\end{equation}
\end{widetext}

\section{Variance of the center of mass}

By way of illustration, we compute the variance of the center of mass $M=N^{-1}\sum_{i=1}^N x_i$ of the Riesz gas. Eq.\ \eqref{varresult} for $p=1$ gives
\begin{equation}
\begin{split}
&{\rm Var}\,M=\frac{C_s\sqrt{\pi }}{2^{s+3} \Gamma (\frac{1}{2}-s/2) \Gamma (2+s/2)},\\
&C_s=\frac{L^{s+2}}{N^2J\beta}\frac{1}{|s|+\delta_{0,s}}.
\end{split}
\label{eqVarM}
\end{equation}
The dimensionless coefficient $C_s$ is $N$-independent if the system is scaled at constant interparticle spacing $\delta x=L/N$ with interaction strength $J\propto N^{s}$.

\begin{figure}[tb]
\centerline{\includegraphics[width=0.8\linewidth]{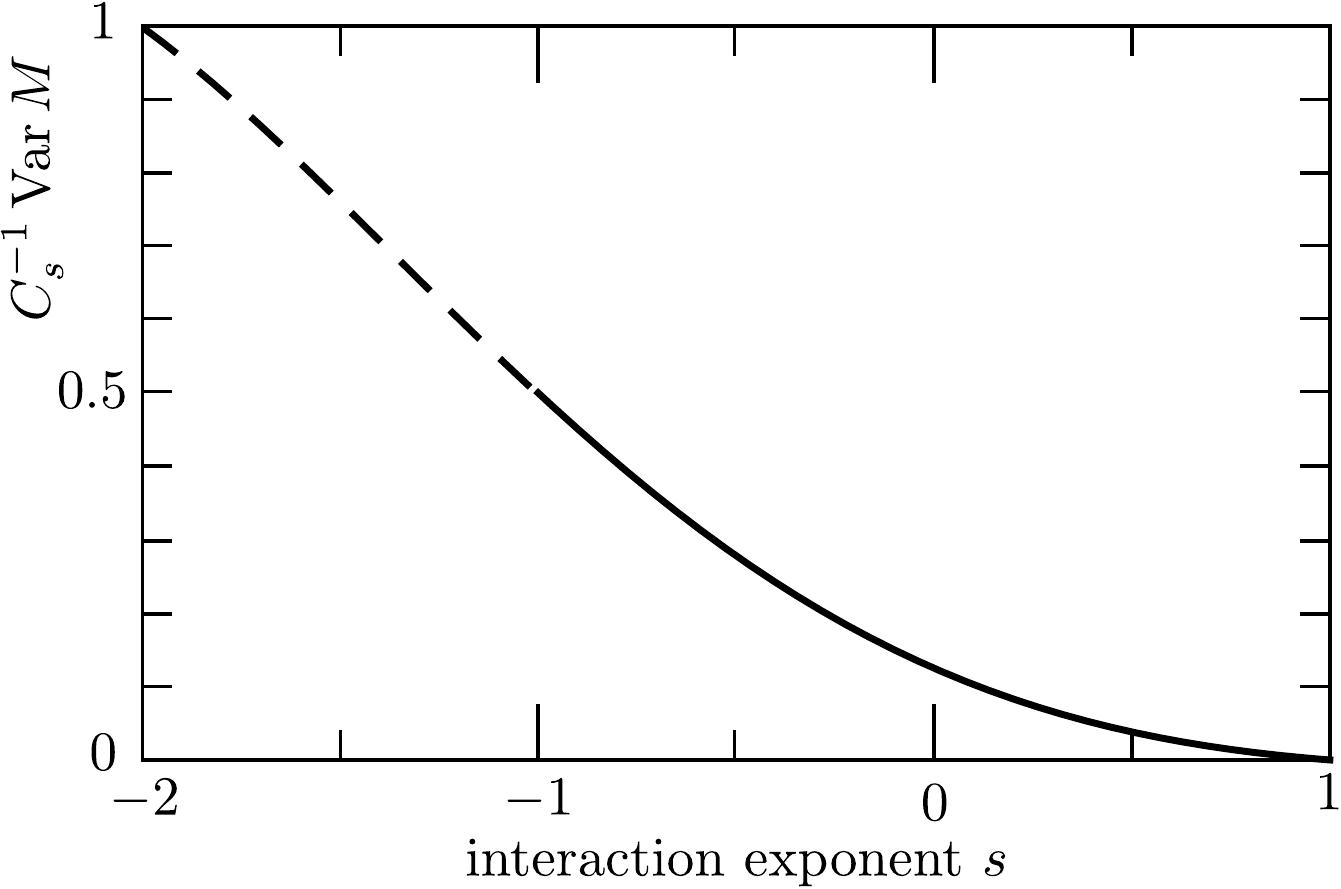}}
\caption{Variance of the center of mass of the Riesz gas, computed from Eq.\ \eqref{eqVarM}. The calculation applies to the interval $|s|<1$, the dashed curve is the analytical continuation of Eq.\ \eqref{eqVarM} to smaller $s$.
}
\label{fig_VarM}
\end{figure}

The dependence of ${\rm Var}\,M$ on the interaction exponent $s$ is plotted in Fig.\ \ref{fig_VarM}. At the upper limit $s\rightarrow 1$ we find $C_s^{-1}{\rm Var}\,M\rightarrow 0$, to leading order in $1/N$. This is consistent with the fact that the repulsion is short-range for $s>1$, so we would expect the positions of the particles to fluctuate independently. The variance of the center of mass (rescaled by $C_s$) would then be of order $1/N$.

The theory applies to the interval $-1<s<1$ of a weakly long-range repulsion. The dashed curve in Fig.\ \ref{fig_VarM} is the analytical continuation of Eq.\ \eqref{eqVarM} to the interval $-2<s<-1$ of a strongly long-range repulsion. We surmise that the formula remains valid in that regime.

\section{Conclusion}

In conclusion, we have computed the pair correlation function of a 1D system of classical particles with a long-range power law repulsion. In the thermodynamic limit (particle number $N$ and system size $L$ to infinity at fixed interparticle spacing $\delta x=L/N$), and upon smoothing over $\delta x$, the pair correlation function becomes a \textit{universal} function of the power law exponent $s\in(-1,1)$ --- independent of the shape of the confining potential $V(x)$ for a given single-interval support $(a,b)$ of the average density. So it does not matter if the Riesz gas is confined to the interval $(0,L)$ by a soft parabolic potential or by a hard-wall confinement --- the density fluctuations are the same even though the average density profile $\rho(x)$ is very different in the two cases \cite{Ket21}.

Our result \eqref{CoVar} for the covariance of a pair of single-particle observables generalizes old results for a logarithmic repulsion \cite{Mehta,Bee96} and a very recent result for a linear repulsion \cite{Fla22}. We rely on a solution of an integral equation that requires $|s|<1$ (weakly long-range regime), but the method of functional derivatives that we have used can be applied also outside of this interval. The independence of the pair correlation function $R(x,y)=-\beta^{-1}\delta \rho(x)/\delta V(y)$ on the shape of the confining potential follows directly from the linearity of the relation between $\rho$ and $V$, so we expect a universal result also for $-2<s<-1$. In the short-range regime $s>1$, in contrast, the $\rho$--$V$ relation is nonlinear \cite{Aga19} and no universal answer is expected.

\acknowledgments
The author receives funding from the European Research Council (Advanced Grant 832256).

\appendix

\section{Sonin inversion formula}
\label{app_Sonin}

We summarize results from Refs.\ \onlinecite{Son54,Sam93,Bul01} on the solution of the singular integral equation
\begin{equation}
{\cal P}\int_a^b dy\,S(y)\frac{\text{sign}(x-y)}{|x-y|^{s+1}}=g'(x),\;\;
x\in(a,b),\;\;|s|<1.
\end{equation}
The homogenous integral equation, with zero on the right-hand-side, has the solution 
\begin{equation}
S_0(x)=[(x-a)(b-x)]^{s_-}.
\end{equation}

The Sonin inversion formula gives two particular solutions to the inhomogeneous integral equation,
\begin{widetext}
\begin{align}
&S_-(x)=-C_1(x-a)^{s_-}\frac{d}{dx}\int_x^b dt\,\frac{(t-x)^{s_+}}{ (t-a)^{s}}\frac{d}{dt}\int_a^t dy\, (y-a)^{s_+}(t-y)^{s_-}g'(y),\\
&S_+(x)=C_1(x-a)^{s_+}\frac{d}{dx}\int_x^b dt\,\frac{(t-x)^{s_-}}{ (t-a)^{s}}\frac{d}{dt}\int_a^t dy\, (y-a)^{s_-}(t-y)^{s_+}g'(y).
\end{align}
\end{widetext}

For the general solution we take either one of these two particular solutions and add it to an arbitrary multiple of the homogeneous solution,
\begin{equation}
S(x)=S_\pm(x)+\text{constant}\times S_0(x),
\end{equation}
where ``constant'' means independent of $x$. Since, by construction, the difference of two particular solutions solves the homogeneous integral equation, it does not matter for the general solution which particular solution we choose. 

In the main text we chose $S_+(x)$. This has the benefit over $S_-(x)$ that the derivatives can be eliminated upon partial integration,
\begin{widetext}
\begin{align}
\int_a^b dx\,f(x)S_+(x)={}&-C_1\int_a^b dx\,\bigl[f'(x)(x-a)+s_+f(x)\bigr](x-a)^{s_-}\nonumber\\
&\times\int_x^b dt\,\frac{(t-x)^{s_-}}{ (t-a)^{s}}\int_a^t dy\, s_+(y-a)^{s_-}(t-y)^{s_-}g'(y).\label{fgnumerics}
\end{align}
The three definite integrals of Eq.\ \eqref{fgnumerics} are in a form that can be evaluated numerically, without the need to take derivatives.
\end{widetext}

\section{Coulomb gas limit}
\label{app_Coulombgas}

The Coulomb gas limit $s=-1$ can be obtained from the general formulas for $|s|<1$ by means of the identities \cite{MO}
\begin{subequations}
\label{identities}
\begin{align}
&\lim_{\epsilon\searrow 0}\int_x^b \frac{\epsilon f(y)}{(y-x)^{1-\epsilon}}\,dy=f(x),\\
&\lim_{\epsilon\searrow 0}\int_a^b \frac{\epsilon f(x)}{[(b-x)(x-a)]^{1-\epsilon}}\,dx=\frac{f(a)+f(b)}{b-a}.
\end{align}
\end{subequations}

We start from the derivative-free representation \eqref{fgnumerics} of the integrals in Eq.\ \eqref{CoVar}, and apply Eq.\ \eqref{identities} first to the integral over $y$,
\begin{align}
I_1(t)&=\lim_{s\searrow -1}\int_a^t dy\, s_+(y-a)^{s_-}(t-y)^{s_-}g'(y)\nonumber\\
&=\frac{g'(t)+g'(a)}{t-a},
\end{align}
then to the integral over $t$,
\begin{align}
I_2(x)&=\lim_{s\searrow -1}\int_x^b dt\,\frac{(t-x)^{s_-}}{ (t-a)^{s}}I_1(t)\nonumber\\
&=\frac{1}{s_+}[g'(x)+g'(a)],
\end{align}
and finally to the integral over $x$,
\begin{align}
I_3={}&\lim_{s\searrow -1}\int_a^b dx\,\bigl[f'(x)(x-a)+s_+f(x)\bigr](x-a)^{s_-}I_2(x)\nonumber\\
={}&\frac{1}{s_+}\int_a^b dx\,f'(x)[g'(x)+g'(a)]+\frac{2}{s_+}f(a)g'(a).
\end{align}
Moreover, since $1/C_2\rightarrow \tfrac{1}{2}s_+(b-a)$ for $s\rightarrow -1$, we have
\begin{equation}
\lim_{s\searrow -1}\frac{1}{C_2}\int_a^b dx\,[(x-a)(b-x)]^{s_-}f(x)=\tfrac{1}{2}[f(b)+f(a)].
\end{equation}

Using also $C_1=\tfrac{1}{2}s_+ +{\cal O}(s+1)^2$ we thus arrive at
\begin{subequations}
\label{limeqs}
\begin{align}
&\lim_{s\searrow -1}C_1\int_a^b dx\,f(x){\cal S}_g(x)\nonumber\\
&\quad=\tfrac{1}{2}\int_a^b dx\,f'(x)[g'(x)+g'(a)]+f(a)g'(a),\\
&\lim_{s\searrow -1}\frac{C_1}{C_2}\left(\int_a^b dx\,{\cal S}_{g}(x)\right)\int_a^b dx\,[(x-a)(b-x)]^{s_-}f(x)\nonumber\\
&\quad=\tfrac{1}{2}g'(a)[f(b)+f(a)].
\end{align}
\end{subequations}
Substitution of Eq.\ \eqref{limeqs} into Eq.\ \eqref{CoVar} gives
\begin{align}
\lim_{s\searrow-1} {\rm CoVar}\,(F,G)=\frac{1}{2J\beta}\int_a^b dx\,f'(x)g'(x),
\end{align}
in accord with Ref.\ \cite{Fla22}.


\begin{thebibliography}{99}
\bibitem{note2} The variational equation \eqref{deltarhodeltaV} does not contain a contribution $\propto\rho(b)\delta b$ from a variation of the end point $b$, for two reasons: Either $\delta b=0$, for a fixed end point, or $\rho(b)=0$, for a free end point. Similarly for the end point $a$.
\end{thebibliography}
\end{document}